\documentclass[9.5pt,journal,compsoc]{IEEEtran}

\ifCLASSOPTIONcompsoc

  \usepackage[nocompress]{cite}
\else
  \usepackage{cite}
\fi
\usepackage{booktabs}
\ifCLASSINFOpdf
   \usepackage[pdftex]{graphicx}
   \DeclareGraphicsExtensions{.pdf,.jpeg,.png}
\else

   \usepackage[dvips]{graphicx}
   \graphicspath{{../images/}}
   \DeclareGraphicsExtensions{.png}
\fi

\usepackage{xcolor}
\usepackage{todonotes}
\usepackage{makecell}
\usepackage{multirow}
\usepackage[normalem]{ulem}

\useunder{\uline}{\ul}{}

\begin{document}

\newcommand{\oya}[1]{{\color{red}#1}}
\newcommand{\hanan}[1]{{\color{blue}#1}}
\newcommand{\etal}{\textit{et al}.}

\title{Automatic Context-Driven Inference of Engagement in HMI: A Survey}

\author{Hanan~Salam,~\IEEEmembership{Member,~IEEE,} Oya Celiktutan,~\IEEEmembership{Member,~IEEE,} Hatice Gunes,~\IEEEmembership{Senior Member,~IEEE,} and 
        Mohamed~Chetouani,~\IEEEmembership{Member,~IEEE,}
\IEEEcompsocitemizethanks{\IEEEcompsocthanksitem H. Salam is with Center of AI \& Robotics (CAIR), SMART Lab, New York University, Abu Dhabi, E-mail: hanan.salam@nyu.edu
\IEEEcompsocthanksitem O. Celiktutan is with the Department of Engineering, King's College London, United Kingdom, WC2R 2LS E-mail: oya.celiktutan@kcl.ac.uk
\IEEEcompsocthanksitem H. Gunes is with the Department of Computer Science and Technology, University of Cambridge, United Kingdom, CB3 0FD, E-mail: hatice.gunes@cl.cam.ac.uk  
\IEEEcompsocthanksitem M. Chetouani is with the Institute of Intelligent Systems and Robotics, CNRS UMR7222, Sorbonne University, Paris, France, E-mail: mohamed.chetouani@sorbonne-universite.fr }}

\markboth{Journal of \LaTeX\ Class Files,~Vol.~13, No.~9, September~2014}%
{Shell \MakeLowercase{\textit{et al.}}: Bare Advanced Demo of IEEEtran.cls for Journals}
\IEEEtitleabstractindextext{%
\begin{abstract}

An integral part of seamless human-human communication is engagement, the process by which two or more participants establish, maintain, and end their perceived connection. 
Therefore, to develop successful human-centered human-machine interaction applications, 
automatic engagement inference is one of the tasks required to achieve engaging interactions between humans and machines, and to make machines attuned to their users, hence enhancing user satisfaction and technology acceptance.
Several factors  contribute to engagement state inference, which include the interaction context and interactants' behaviours and identity. Indeed, engagement is a multi-faceted and multi-modal construct that requires high accuracy in the analysis and interpretation of contextual, verbal and non-verbal cues.
Thus, the development of an automated and intelligent system that accomplishes this task has been proven to be challenging so far. 
This paper presents a comprehensive survey on previous work in engagement inference for human-machine interaction, entailing interdisciplinary definition, engagement components and factors, publicly available datasets, ground truth assessment, and most commonly used features and methods, serving as a guide for the development of future human-machine interaction interfaces with reliable context-aware engagement inference capability. An in-depth review across embodied and disembodied interaction modes, 
and an emphasis on the interaction context of which engagement perception modules are integrated sets apart the presented survey from existing surveys.

\end{abstract}

\begin{IEEEkeywords}
Engagement Detection, Human-Machine Interaction, Socially Intelligent Systems.
\end{IEEEkeywords}}

\maketitle

\IEEEdisplaynontitleabstractindextext

\IEEEpeerreviewmaketitle
\section{Introduction}
\label{sec:intro}
The field of human-machine interaction (HMI) is rapidly developing to address various societal challenges. Human interactions with machines can take different forms, depending on the scenario, machine (dis)embodiment (referred to as interaction mode hereafter), and interaction goal. Examples include delivering remote education \cite{ikedinachi2019artificial}, enhancing mental well-being \cite{thieme2020machine}, and supporting elderly individuals \cite{broekens2009assistive}. 
 The success of such applications highly depends on users' satisfaction, trust and technology acceptance; therefore, it is becoming increasingly desirable that human-machine interaction systems develop social intelligence and become attuned to their users through the effective use of multimodal communication channels, ultimately leading to the maximization  of the targeted interaction's outcomes \cite{tatarian2021does}. 

One key component of social intelligence is unarguably engagement \cite{goffman1963behavior,o2008user,Sidner2002}. Engagement is a complex multi-modal and multi-faceted social phenomenon that requires the perception and recognition of social signals and their interpretation at a higher level for social behaviour regulation. 
In the past decade, there has been a significant body of work that aims to develop engagement inference models and machine behaviour adaption mechanisms in various human-machine interaction contexts \cite{del2020you,el2019learning}.  From the perspective of disembodied interaction (machine interface without embodiment, e.g. human-computer interaction), for example in the context of HMI for learning, it is important to design engaging learning systems that have the capability of detecting the user's engagement state and adapting to it, allowing the user to acquire the learning outcomes objectives \cite{dewan2019engagement,nasir2022many}. Within the context of game entertainment, designing engaging games is essential for making the user's experience enjoyable and preventing withdrawal \cite{rapp2022time}. From the perspective of embodied interaction (machine interface with physical or virtual embodiment, e.g. human-agent interaction and human-robot interaction), engagement is an essential rubric, which allows a smooth and natural interaction between the user and the robot/agent, and can contribute to achieving  effective long-term interactions that go beyond the novelty effect. Across different HMI contexts, achieving engaging human-machine interactions requires that the machine is able to 1) interpret human's engagement from the observation of their multimodal  cues~\cite{sidner2005explorations} and 2) express its engagement in an appropriate manner beyond on-off interactions. 

Lately there has been an increasing trend towards integrating contextual information in social signal processing and affective computing research \cite{castellano2014context,salam2015engagement}. During an interaction, any information that allows the characterization of an entity's situation can be considered as context, provided that the entity is an individual, a location, or any object relevant to the human-machine interaction, including the user and the machine \cite{abowd1999towards}. Context is used intuitively by humans in social interactions to act and react properly, as well as to correctly infer the others' state of mind \cite{abowd1999towards}. In particular, humans  might manifest their engagement state in different ways that largely depends on context \cite{witchel2014does}. The user's mental, emotional, and behavioural states associated with their  engagement state was also found to vary with the interaction context~\cite{salam2015multi}. Consequently, improving the machine's access to  context information would increase its social intelligence skills and promote more accurate, adaptive, and engaging user experience~\cite{abowd1999towards}.

The importance of integrating context in the design of engagement inference systems has been loosely underlined in the literature through the use of various contextual cues as input to automatic engagement inference models \cite{castellano2012detecting}. 
Despite the global recognition of the importance of context-aware  engagement modeling and inference by the community, however, the literature lacks a systematic context-driven overview on the topic.
In this paper, we present an in-depth overview of the engagement modelling, detection, and recognition approaches across different interaction modes (i.e. disembodied  and embodied interaction).  We put a special emphasis on the importance of context for modelling engagement and for the development of context-aware, accurate and adaptive engagement perception algorithms.  The current engagement survey is context-driven in the sense that it discusses engagement definitions across different interaction modes, outlines different contextual factors that have an effect on human-machine engagement (e.g. interaction mode and scenario, and personal factors), and reviews contextual features used in engagement inference models in the literature.

There has been a couple of previous efforts that reviewed the definition of engagement~\cite{glas2015definitions} and its implications in human-agent interaction~\cite{o2008user}. There is also a recent survey by Oertel~\etal~\cite{10.3389/frobt.2020.00092}, which reviewed the definition of engagement and how it differs across different interaction settings (i.e., real-world versus laboratory, short-term versus long-term, social versus task oriented) and user profiles (e.g., adults versus children). However, in their survey, the emphasis is more on the behaviour adaptation strategies and the review of the engagement perception methods does not go beyond the deep learning approaches. In \cite{lytridis2019measuring}, a survey on engagement level recognition in child-robot interaction was also presented. However, the survey's scope was limited to education and therapeutic settings. 
To the best of our knowledge, this paper is the first comprehensive context-driven survey of automatic engagement inference, starting from the definition of engagement to the design of the full detection pipeline including data acquisition, feature extraction, and inference.

This  review will serve as a guide for researchers interested in the topic of engagement to acquire a holistic understanding of the concept. Specifically, the emphasis on the contextual factors of engagement, and how engagement was defined and detected across different contexts in the literature will inform context-aware artificially intelligent systems. Consequently, this will allow the design of  human-machine interactions with increased  usability, accuracy, and efficiency in real-time settings, leading to improved user experience.

\section{Context-Driven Engagement Definition}
\label{sec:def}

In order to build effective systems for engagement inference, it is essential to establish a clear and precise definition of the notion. Engagement is a complex construct composed of various components or factors, which were covered across various interaction modes, namely, human-human interaction (HHI), human-computer interaction (HCI), human-agent interaction (HAI) and human-robot interaction (HRI). Different concepts (e.g.  attention, involvement, interest, immersion, rapport) were related to engagement and sometimes even used interchangeably in the literature \cite{glas2015definitions}.  

From the perspective of the user, engagement is often seen to be composed of three factors: emotional, cognitive, and behavioural. 
Depending on the context, some factors can be predominant with respect to the others~\cite{salam2015multi}. For instance, cognitive factors such as concentration might be more predominant in a learning context, compared to a purely social context. It is crucial to understand how the notion of engagement changes based on the different context categories as they appeared in the different studies. We distinguish between three context types that influence the user's engagement state, and process in their interaction with intelligent systems, namely, (1) the interaction mode (i.e., embodied vs. disembodied), (2) the interaction scenario (e.g., competitive vs. collaborative), and (3) personal factors (e.g., personality and gender). In this section, we summarise the most commonly used definitions of engagement and associated attributes. We investigate how engagement was related to different concepts across different contexts. We underline how these change based on the context. Table \ref{tab:definitions} summarizes the commonly used definitions across the different modes of interaction.

\begin{table*}[htbp]
\scriptsize
\label{tab:definitions}
\centering
\caption{Summary of the most commonly used definitions across the different modes of interaction: HHI, HCI, HAI, and HRI. Interaction Mode (IM).}
\begin{tabular}{lm{15cm}l}
\toprule
\textbf{IM}                      & \textbf{Definition}    & \textbf{Paper}                                      \\
\midrule
\textbf{HHI}                                            & Engagement occurs when people gather closely together and openly cooperate to sustain a single focus of attention, typically by taking turns at talking.                                                                                                                                                                                                                                                                       &  \cite{goffman1963behavior}         \\
\midrule
\textbf{HCI}                                            & Engagement with technology is a measure of the quality of user experience.   &  \cite{lehmann2012models,o2008user} \\
\cmidrule{2-3}
& A connection that exists at any point of time and possibly over time between a user and a resource.                                                                                   &                                                     \\
                                               & The cognitive, affective, and behavioural state of interaction that makes the user want to be there.  & \cite{obrien2010development}       \\
                                               \cmidrule{2-3}
                         & Engagement in online learning is a construct that encompasses student's behaviours and involvement in consistent engagement with resources or activities within the online environment, with the end-goal of achieving learning.    &  \cite{brown2022conceptual}         \\
                         \cmidrule{2-3}
 & Engagement in computer supported collaborative learning is with-me-ness which measures how much are the students with the instructor.             &  \cite{sharma2014me}    \\
\midrule
\textbf{HAI}                                            & An emotional state linked to the participant's goal of receiving and elaborating new and potentially useful knowledge.  &  \cite{peters2005model}             \\
\cmidrule{2-3}
                                               & \textit{Empathic engagement} is fostering of emotional involvement intending to create a coherent cognitive and emotional experience which results in empathic relations.  &  \cite{hall2005achieving}           \\
                                               \cmidrule{2-3}
                                               & The value that a participant in an interaction attributes to the goal of being together with the other participant(s) and continuing interaction.    &  \cite{poggi2007mind}               \\
                                              
                                               \midrule
\textbf{HRI}                                            & The process by which two (or more) participants establish, maintain and end their perceived connection. This process includes initial contact, negotiating a collaboration, checking that the other is still taking part in the interaction, evaluating, staying involved, and deciding when to end connection.  &  \cite{Sidner2002}                  \\
\cmidrule{2-3}
                                               & The process of subsuming the joint, coordinated activities by which participants initiate, maintain, join, abandon, suspend, resume, or terminate an interaction.   &  \cite{Bohus2009}                   \\
                                               \cmidrule{2-3}
                                               & Magnitude of an intrinsically motivated behaviour that is initiated by an organism to reach a specific goal &  \cite{drejing2015engagement}       \\
                                               \cmidrule{2-3}
                                               & \textit{Task engagement} where there is a task and the participant starts to enjoy the task they are doing, \textit{social engagement} which considers being engaged with another party when there is no task included, and \textit{social-task engagement} which includes interaction with another (e.g., robot) where both cooperate with each other to perform a task. &  \cite{corrigan2013social}          \\
                                               \cmidrule{2-3}
                                               & \textit{Productive Engagement} is defined as the level of engagement that maximizes learning. 
                                               &  \cite{nasir2021if}   \\
\bottomrule 
\end{tabular}
\end{table*}

\subsection{Interaction Mode: Embodied vs. Disembodied}

In the sequel, we review widely used definitions of engagement for human-human interaction, disembodied human-machine interaction (e.g., mobile devices and web applications), and embodied human-machine interaction (e.g., agents, and robots) settings. 

\begin{figure}
    \centering
    \includegraphics[width=0.5\textwidth]{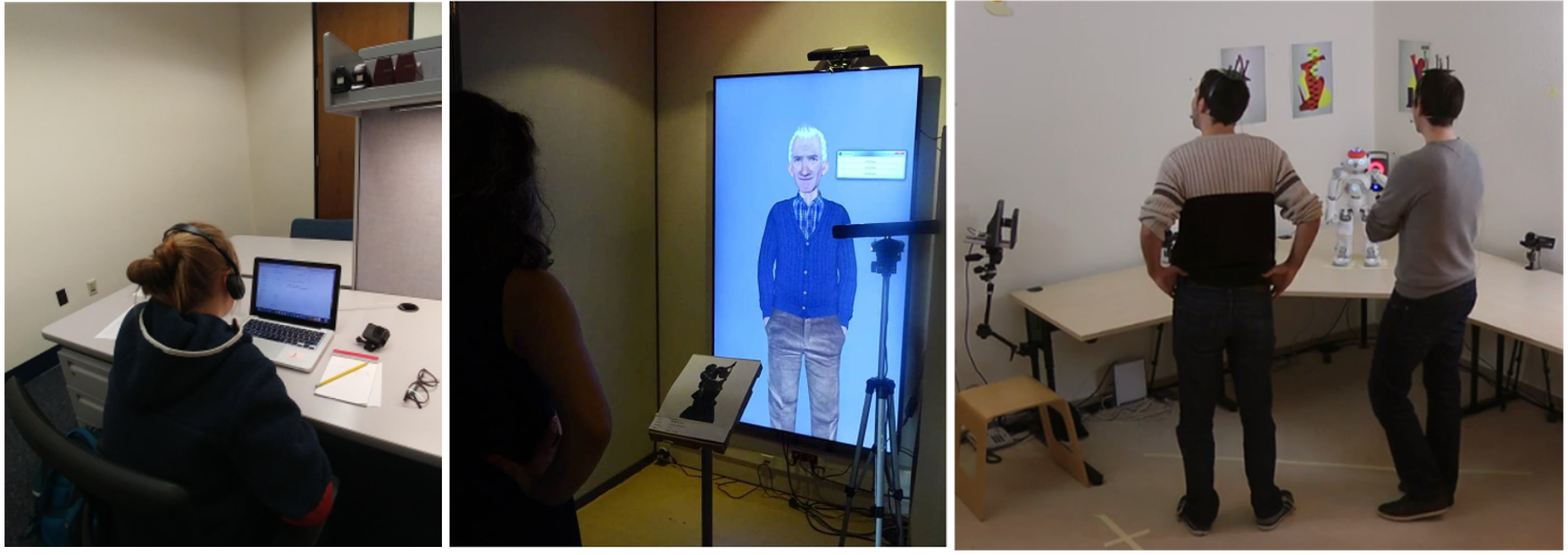}
    \caption{Examples of disembodied~\cite{delgado2021student} and embodied (HAI \cite{glas2015user}, HRI \cite{salam2015engagement}) interaction scenarios.}
    \label{fig:ex_interactions}
\end{figure}
\subsubsection{Engagement in Human-Human Interaction }
 
The notion of engagement in human-human interaction (HHI) was addressed in social sciences by Goffman~\cite{goffman1963behavior} who differentiates between \textit{unfocused interaction} and \textit{focused interaction}. Unfocused interaction, a pre-requisite of engagement, is concerned with what can be communicated between people due to their co-presence in the same social situation (e.g. glancing to glean information about the other). On the other hand, focused interaction (face engagements/encounter) takes place when
there is \textit{cooperation} between individuals gathering closely together to \textit{sustain a single focus of attention}, commonly via turn-taking and conversing.

In HHI, several researchers investigated engagement in the context of learning. Learner's engagement was linked to emotional (valence, value, interest), cognitive (motivation, effort, strategy), and behavioural (participation, persistence) components \cite{zyngier2008re} \cite{fredricks2004school} \cite{Peters09} \cite{Monkaresi2014} \cite{whitehill2014faces}. 
Another view of learner's engagement in school setting relates engagement to behavioural, academic, cognitive, and psychological dimensions  \cite{anderson2004check}.
For instance, behavioural engagement in a class refers to class attendance, concentrating on tasks, listening and following the teacher's directions. A cognitive component (concentration) is apparent in what is described as behavioural in \cite{anderson2004check}.
\emph{Emotional engagement} is the emotional attitude concerning the learning task. A student may have high behavioural engagement (a student obtaining high grades on exams) and low emotional engagement (but is bored). 
\emph{Cognitive engagement} is related to learning with cognitive abilities such as memory, attention, or strategy. 
Another work considered student's engagement as the linear sum of the student's perceived focused attention, percieved involvement in the task, and the endurability (perception of the experience as worthwhile) \cite{vail2014predicting}. 
Few studies discussed that a learner's engagement state can be modelled by the learner's affective state \cite{Papadopoulos2013}. Specifically, engagement was directly related to the state of flow \cite{graesser2006detection}, which can be reached during periods of full engagement (e.g. when improving or enjoying the learning activity), whereas disengagement can be depicted during period of lack of enjoyment or non-advancement in the learning activity. Negative affective states such as boredom, frustration, and confusion are connected to disengagement \cite{graesser2006detection}. 
 In addition to enjoyment, factors of challenge and being in a zone of proximal development (distance between what a learner can do with support and without support) were considered for student's engagement \cite{brown1998creating}. 
The study of Pekrun~\etal~\cite{pekrun1992impact} proposed that learning and cognition are highly affected by affect, which has an impact on motivation, attention, and strategy use. 

The definitions of engagement in HHI focus on the co-presence of interaction partners, cooperation on mutual activities, sustaining a single focus of attention, and establishing and maintaining a connection. The interaction between humans is an embodied one, and the embodiment is similar. Humans as interaction partners are of similar nature, and the interaction is governed by various factors such as the (in)existence of a history between the individuals, their social roles, the goal of the encounter, their knowledge of each others, their rapport, their similarities and differences, their characteristics and backgrounds, etc.

\subsubsection{Engagement in Disembodied HMI} 
Disembodied HMI (also referred to as HCI) can take several forms, such as interacting with a computer application, web searching, online shopping, webcasting, learning, and gaming among others. In such interaction, the focus is primarily on the task. 
 While some form of social interaction might exist, the absence of embodiment, limits the machine's expression of some forms of social intelligence (e.g. through non-verbal signals like gestures and emotions), and consequently, the expectation of the user from the system in this regard. Certain social channels (e.g interactive pop-up messages in learning contexts) can be used in such contexts to increase the human's engagement, but this depends on the machine's design and social expressivity capabilities.
 However, the design of engaging human-computer interactions,  even with limited social intelligence expressivity,  is of utmost importance, and serves in attaining the system's end-goal, ensuring its usability and long-term usage. The factors pertaining to engagement in HCI depend on the system's characteristics, social expressivity capabilities, and the interaction's end goal, among others.  

Engagement with technology is regarded as a measure of the quality of user experience \cite{lehmann2012models,o2008user}, a connection between a user and a resource that can exist at any instant and even in the long term \cite{attfield2011towards}. Engagement with a computer application was referred to by Obrien~\etal~\cite{obrien2010development} as the cognitive, affective, and behavioural state of interaction that ``makes the user want to be there''. It is a process comprised of four distinct stages including point of engagement, period of sustained engagement, disengagement, and re-engagement. It is characterized by attributes that concern the user, system, and their interaction. These include challenge, positive affect, endurability, aesthetic and sensory appeal, attention, feedback, variety/novelty, interactivity, and perceived user control \cite{o2008user}. Reputation, trust and expectation, and user context were also underlined as characteristics of engagement in the context of web-applications \cite{attfield2011towards}. In practice, engagement characteristics were narrowed down in some HCI contexts to cognitive components, e.g. attention \cite{asteriadis2009feature,Belle2012}, or affective components, e.g. frustration \cite{sun2014nonintrusive}.

Engagement in the context of online learning has been widely studied. For instance,  the concept of with-me-ness for computer supported collaborative learning was introduced in \cite{sharma2014me} in an attempt to assess ``how much are the students with the instructor?''. Two components of with-me-ness are distinguished: perceptual \-- learner engages with what the instructor is referring to via deictic gestures (e.g. pointing), and  conceptual \-- learner engages with what the instructor is referring to verbally. 
In \cite{brown2022conceptual}, engagement in online learning is regarded as a construct that encompasses student's behaviours and involvement in consistent engagement with resources or activities within the online environment, with the end-goal of achieving learning. Concerning factors pertaining to engagement in online learning were referred to as social, cognitive, behavioural, collaborative and emotional elements \cite{redmond2018online}. The student's access to learning material was underlined as a factor of student's engagement in \cite{soffer2019students}. A longitudinal study aiming at understanding online student engagement over a semester \cite{muir2019chronicling} demonstrated a  dynamic and fluctuating nature of student engagement, which is affected by factors of  assessment, course units workload, course units content, lecturer presence and behaviour, and work/life commitments. 

The definitions of engagement in disembodied HMI focus on the user-system interaction end-goal. While some factors are common to HHI, such as the user's cognitive, emotional, and behavioral factors, other factors pertaining to the system (e.g. aesthetics, reputation, perceived user control) or to the interaction context (e.g. course units content and workload in online learning) are more relevant to HCI.

\subsubsection{Engagement in Embodied HMI}
Embodiment in HMI usually takes two forms: virtual (embodied conversational agents), and physical (robots). Embodiment results in higher social intelligence expressivity capabilities in the system and interaction as compared to disembodied interaction. Interaction with virtual agents is often conversational, which adds more weight to the social aspect of engagement. On the other hand, physical embodiment increases the domain of applications of the machine. Activities that require physical capacities (e.g. waitressing or care-giving) as well as physical collaborative activities (e.g robot-mediated collaborative learning) become possible with physical embodiment. 

\textbf{HAI.} In the literature of virtual embodied agents, some definitions of engagement were associated with the goal of acquiring knowledge, with an emphasis on the emotional dimension and empathy. For instance, in the context of a conversational scenario, Peters \etal \cite{peters2005model} defined engagement as ``an emotional state linked to the participant's goal of receiving and elaborating new and potentially useful knowledge''. They presented  engagement as the direct consequence of interest and and attention. 
The notion of empathic engagement was referred to by
\cite{hall2005achieving} as ``fostering of emotional involvement intending to create a coherent cognitive and emotional experience which results in empathic relations''. According to Glas and Pelachaud \cite{glas2015definitions}, the definition of Poggi \cite{poggi2007mind} of engagement as ``the value that a participant in an interaction attributes to the goal of being together with the other participant(s) and continuing interaction'' is the most suitable definition for engagement in embodied HMI as it resonates with making interactions with virtual agents as believable and human-like as possible.

\textbf{HRI.} The history of engagement in the field of HRI dates back to the year 2002. In~\cite{Sidner2002}, Sidner and Dzikovska presented a robot endowed with basic engaging capabilities including the capacity to initiate, maintain, and end a conversation with a human. According to Sidner and Dzikovska, engagement is defined as ``the process by which two (or more) participants establish, maintain and end their perceived connection. This process includes initial contact, negotiating a collaboration, checking that other is still taking part in interaction, evaluating, staying involved, and deciding when to end connection.'' In their work \cite{Sidner2002}, engagement was studied in the context of hosting activities, namely, a class of \textit{collaborative activities} where an agent provides services to the human in a certain context (e.g. information, entertainment, education), such that the human may be requested by the agent to perform some actions,  necessary for the fulfilment of such services.
They underlined attention as a direct correlate of engagement by demonstrating that, during an interaction, a robot performing ``engagement gestures'' (e.g. following the user's face) would lead to an increase of user's attention, and consequently, their engagement. 

The definition of Sidner and Dzikovska is among the most widely used definitions in the area of HRI. For example, \cite{rich2010recognizing}, \cite{bohus2014managing}, and \cite{castellano2012detecting} adopted this definition of engagement, particularly, for describing emotional interaction level and social bonding established in child-robot interaction. Other definitions widely used in the HRI literature were proposed by Bohus \etal~\cite{Bohus2009} and Peters~\cite{Peters2005}. According to Bohus \etal \cite{Bohus2009}, engagement is ``the process of subsuming the joint, coordinated activities by which participants initiate, maintain, join, abandon, suspend, resume, or terminate an interaction.'' 
Various HRI studies employed the conceptualisation of Bohus~\etal's engagement definition in their studies \cite{foster2013can,Klotz2011,Yun2012}. The definition of Poggi~\cite{poggi2007mind} was adopted by Peters~\cite{Peters2005} and used in the context of child-robot interactions by Castellano \etal~\cite{Castellano2009} and Sanghvi \etal~\cite{sanghvi2011automatic} among others.  Another definition that builds upon motivation theory instead of cognitive and emotional constructs is that of Drejing~\etal~\cite{drejing2015engagement} who suggests that engagement can be defined as the
magnitude of an intrinsically motivated behaviour initiated by an individual to reach a specific goal. 

With physical embodiment, the interaction components can vary between social, task, and social-task. Consequently, an interesting view on engagement, in this regard, is that of Corrigan \etal~\cite{corrigan2013social} who introduced a context-dependent engagement definition, in terms of task, social and social-task contexts. \textit{Task engagement} corresponds to scenarios where a participant is performing a task, and enjoys it. \textit{Social engagement} involves engagement  with an interaction partner in a social interaction, with an absence of a task. \textit{Social-task engagement} involves a cooperation with another to perform a task. Some studies such as that of~\cite{lemaignanreal2016} focused on social-task engagement, by concentrating on the degree of engagement of the user with the robot during a collaborative task. Consequently, a metric of engagement was defined as the normalized fraction of time an interaction party directs their attention to the attention target the robot expects for the current task (or sub-task).
In the context of collaborative learning, the concept of Productive Engagement (PE) was introduced in \cite{nasir2021if} with the aim to conceptualize \textit{engagement that is conducive to learning}. 
PE is defined as the level of engagement that maximizes learning and is composed of  social and task engagement. In contrast to existing work in engagement conceptualization, the authors argue that being overly engaged can result in decreased learning outcomes.

Similar to disembodied HMI, cognitive and emotional elements of attention and valence of feeling were emphasised as components of engagement \cite{castellano2014context}. According to Corrigan \cite{corrigan2013social}, engagement is characterized by elements of 
participation, commitment, concentration, involution and immersion. Causation elements constitute internal states or desires like intrigue, curiosity, amazement, interest, concern, or wonder. 
Furthermore, engagement may evoke more emotional aspects of awareness, states of pleasure or arousal, thus justifying the initial investment in engagement.

Engagement and its underlying factors in embodied HMI can be seen as a middle-way construct, borrowing aspects from engagement in HHI and disembodied HMI. On the one hand, embodiment provides presence, and humans are less likely to ignore the system, and consequently the expectations in terms of social interaction naturalness and human-likeness. On the other hand, the focus on the task and its differentiation from the social aspect of the interaction is  apparent in the embodied HMI engagement literature. 

\subsection{Interaction Scenario}

Interaction scenario describes the interaction between the machine (embodied or disembodied) and the human. Different interaction scenarios might trigger different cognitive, emotional, and behavioural user states, indicative of their engagement state \cite{salam2015multi}. Based on existing literature in HMI, we differentiate between $7$ interaction scenarios: (1) \textit{Purely social}, (2) \textit{Informative}, (3) \textit{Educative},  (4) \textit{Competitive}, (5) \textit{Collaborative}, (6) \textit{Negotiation},  and (7) \textit{Guide-and-follow}. Some scenarios are possible for all interaction modes (e.g. competitive, informative), while others are restricted to embodied interaction (e.g. collaboration, negotiation, social). 

\textbf{Purely social} \cite{michalowski2006spatial} context is a social context that does not involve performing a task. In such context, social interaction may include greetings, self-introductions, or informal talking, etc. In conversational HCI context, temporal characteristics \cite{yu2004detecting} composed of the user's past engagement state (temporal continuity), their current emotional state, and the other participants' engagement states were considered. Cognitive factors including attention and interest in the conversation were related to engagement.  

\textbf{Informative} \cite{michalowski2006spatial,xu2013designing} context entails transmitting general information that does not fall in the category of educating, e.g. giving navigation directions to reach a certain location.

\textbf{Educative} \cite{kapoor2005multimodal} context describes a form of learning  which entails the transfer of knowledge or skills from an educator to a learner. Student engagement within the context of technology supported learning has been widely studied in the literature~\cite{KamathEtAlWACV16,daisee_dataset,kaur2018prediction} from physiological signals or videos recorded during Massive Open Online Courses (MOOCs), or data collected in the class~\cite{alyuz2016towards,cukurova2020modelling,KasparovaEtAlICMI20Companion}. 

\textbf{Competitive} \cite{castellano2012detecting, jayagopi2013given,sheikhi2013context} context is characterized by elements of skills or knowledge testing or a form of competition over a certain profit (e.g. quiz, non-collaborative game, etc.). In such context, states of concentration and reflection might be triggered.

\textbf{Collaborative} \cite{sidner2003engagement} context involves a form of collaboration to achieve a pre-defined task. Such scenario is more prevalent in HRI, since physical embodiment allows a wider range of collaborative activities. 

\textbf{Negotiation} \cite{nouri2013prediction} context involves the adoption of several strategies to achieve goals. Various parties confer and reach an agreement.

\textbf{Guide-and-follow} \cite{sidner2004look} context is concerned with  cases involving a form of guidance to accomplish a specific task, where one party leads while the other follows the directions.

\subsection{Personal Factors}
\label{sec:personalfactors}
Personal factors such as the user's gender, culture, age, ethnicity,  personality, or if the user has a certain pathology were underlined in the literature as intrinsic factors affecting the engagement state of an individual \cite{park2012law}. 
From the system's perspective, specifically in embodied HMI, the system's personal factors (e.g. gender, personality) were also found to directly affect the user's engagement state. Other personal factors such as age and ethnicity are not explored in the literature of engagement inference. This might be related to the fact that most studies in engagement restrict their studies to specific age groups (e.g. adults or children), and it is rare to find studies with datasets that include different age groups or different ethnicities. 

\textbf{Gender}. 
The effect of gender on user's engagement was discussed in some studies \cite{park2011effects,sidner2004look}. For example, the HRI study of \cite{park2011effects} found that most participants preferred interacting with robots of the opposite sex, with a stronger gender effect when the participant was a male, and the robot is a female. This suggests that designing human-machine interactions that adapts the gender of the machine to that of the user would result in higher user engagement. Similarly, also in an HRI context, it was found that female and male users engage differently with robots \cite{sidner2004look}.
Personalized models such as gender-specific models improved the accuracy of inference compared to general models in HCI scenarios \cite{Monkaresi2014}.\\

\textbf{Culture}. Differences in social behaviour among different societies and cultures has been thoroughly underlined in the literature
\cite{triandis1994culture}. For instance, studies on emotion recognition have reported higher accuracy when cultural factors were taken into consideration \cite{elfenbein2002universality}, encouraging the adoption of a culture-sensitive approach in the assessment of emotions \cite{mesquita2003cultural}, and consequently emotion-dependent constructs such as engagement. Similarly, in HRI, an analysis of children engagement have revealed differences in engagement displays across different cultural backgrounds \cite{rudovic2017measuring}, which have been taken into account in the computational models of engagement \cite{rudovic2018culturenet}. A study on proxemics (a relevant cue of engagement) in HRI, particularly on robot approaching groups of people, have also shown different preferences in proxemics behaviour among different cultures \cite{joosse2014cultural}. In HCI, it was found that between- and within-country cultural differences have an impact on digital consumer engagement and engagement with online marketing material \cite{joosse2014cultural}. In the context of learners engagement, cultural factors were found to be correlated with organisational, technological, and pedagogical components of online learning \cite{hannon2007cultural}. \\

\textbf{Personality}. The effect of personality on the engagement state is also evident in the literature. For instance, in contrast to gender, previous studies found that human interactions with a robot having the similar personality traits were perceived as more comfortable compared to interactions with a robot having  different personality traits \cite{park2012law}. 
In a triadic HRI study, results showed a significant correlation between the perceived enjoyment with an extroverted robot and the participants'  agreeableness and extroversion traits \cite{celiktutan2015computational}.
The effect of the user's personality regardless of the robot's personality on the user's engagement state was also investigated. 
Findings indicate that higher extroversion scores were correlated with longer interactions with robots \cite{Ivaldi2015}.
High  conscientiousness scores were associated with higher expression of attentiveness and responsiveness in interaction \cite{cuperman2009big}. On the other hand, individuals scoring high on the agreeableness dimension reported higher enjoyment in interactions compared to others. 
In learning contexts, it was found that extroverted and introverted students   exhibit different behaviours to indicate the same cognitive and affective states \cite{vail2014predicting}.

\textbf{Pathology.}
The presence of certain pathologies can alter the way an individual behaves. For example, pathologies that have an effect on social behaviour include Autism Spectrum Disorder (ASD), Attention Deficit Hyperactivity Disorder (ADHD), Major Depression Disorder (MDD), Bipolar Disorder (BP), among others. For instance, individuals with ASD are characterized by deficiencies in demonstrating proper social cues in social interaction contexts \cite{american2013diagnostic}. Individuals with ADHD may exhibit an increased quantity of movement, in addition to an impaired capacity of sustaining attention on tasks. The characteristic social behaviours of individuals with such pathologies compromises a generic engagement model ability to accurately infer the human's engagement state.  The fact that individuals attained with such pathologies exhibit unusually diverse styles in the expression of their affective-cognitive states makes the inference task even more challenging \cite{rudovic2018personalized}. Consequently, integrating the pathology information or clinical assessments in engagement inference models might better inform the decisions of such models. Except for few approaches, the use of such information is scarce in the literature. For instance, clinical assessment in addition to  culture, gender and individual traits information were proposed in \cite{rudovic2018personalized} to condition autoencoders   for the task of inferring   child's engagement level continuously in time, in the context of robot-assisted autism therapy.

\section{Automatic Inference of Engagement}
\label{sec:automaticDet}
In this section, we present a detailed summary of the existing approaches to automatic engagement recognition, including an overview of data acquisition for engagement inference, the multimodal behavioural cues commonly used as features for engagement in the literature and the employed machine learning approaches for the task, both traditional and modern solutions (e.g., deep learning).

\subsection{Data Acquisition}
Usually data is collected using unimodal or multimodal sensors such as microphones, cameras, 3D sensors (e.g. Kinect\textsuperscript{\textcopyright}), and physiological sensors. However, the choice of modalities depends on the context of application. 
Collected samples are then given ground truth labels reflecting the perceived or reported engagement state of the user. Engagement ground truth assessment largely depends on the context in which the engagement is being measured. Compared to  emotion data annotation where categorical and dimensional scales are commonly used to assess emotions ground truth \cite{constantine2012survey}, there are no common scales used to collect engagement ground truth. Whether to treat engagement as a process, a discrete or a continuous value, or to concentrate on a specific component of the construct (e.g. behavioural vs. cognitive engagement) is highly dependent on the interaction context and end-goal. Moreover, there is no agreement on the optimal time scale for engagement annotation, e.g. frame-level or segment-level (cf. Section \ref{sec:temporaldynamics}).
Indeed, in the literature, there is no unified strategy for user engagement state annotation. 
In the following, we give an overview of the publicly available engagement datasets (Section~\ref{sec:datasets}) as well as common annotation strategies to obtain ground truth data (Section~\ref{sec:annotation}). We finalise with reviewing the problem formulation for engagement and the definition of categories that commonly appear in the state-of-the-art (Section \ref{sec:catview}).

\subsubsection{Publicly Available Datasets}
\label{sec:datasets}

Since engagement is a context-dependent construct and a relatively recent subject in human-machine interaction, there are only a few publicly available datasets in the literature, which provide engagement annotations. These datasets are given in Table~\ref{tab:datasets} and are introduced based on the interaction mode below.  

\begin{table}[htbp]
\scriptsize
  \centering
  \caption{Overview of publicly available datasets in engagement inference. Modalities: Audio (A), Video (V), Physiological (P), Log -- data that captures interaction with the setup (L).}
  \label{tab:datasets}
 \tabcolsep=0.1cm
  \begin{tabular}{llllm{2cm}l}
  \toprule
 \textbf{Mode}  &\textbf{Dataset} &  \textbf{Modality}  & \textbf{$\#$ S} & \textbf{Context} & \textbf{Papers}\\
  \bottomrule
  \multirow{1}{*}{\textbf{HHI}}&\makecell[l]{RECOLA~\cite{ringeval2013introducing}\\($2013$)}  & V, A, P & 46 & Collaborative& $--$\\  [-3pt] 
  \midrule   
 \multirow{6}{*}{\textbf{HCI}} &\makecell[l]{DAiSEE \cite{gupta2016daisee}\\($2016$)} &  V & 112  & Educative &\cite{abedi2021affect}\\ [-6pt] 
 &\makecell[l]{HBCU \cite{whitehill2014faces}\\($2014$)} &  V & 34 &  Educative   & $--$\\ [-6pt] 
&\makecell[l]{in-the-wild\cite{kaur2018prediction}\\($2018$)}&  V & 78 & Educative &\cite{niu2018automatic}\\ [-3pt] 
\midrule   
 \multirow{10}{*}{\textbf{HRI}} &\makecell[l]{MHHRI \cite{celiktutan2017multimodal}\\($2017$)}& V, A, P&18 &  Social&\cite{salam2016fully}\\ [-6pt] 
&\makecell[l]{PE-HRI  \cite{nasir2020pe}\\($2020$)} &V, A, L&68 &Educative&\cite{nasir2021if}\\ [-6pt] 
&\makecell[l]{PE-HRI-temporal \cite{nassir2021dataset}\\($2021$)}& V, A, L&68&Educative&\cite{nasir2022many}\\ [-6pt] 
&\makecell[l]{UE-HRI \cite{ben2017ue}\\($2017$)}  &V, A&54 & Social &\cite{liu2018predicting,ben2019fly}\\
\bottomrule
  \end{tabular}
\end{table}

\textbf{HHI}. The Remote Collaborative and Affective Interactions (RECOLA) database 
\cite{ringeval2013introducing} provides engagement labels in addition to set of affective and social behaviour annotations including arousal, valence, agreement, dominance, performance, and rapport in a mediated interaction context. The participants in this corpus were recorded remotely in dyads during a video conference while completing a collaborative task (the survival task). In addition to video, the corpus includes audio and physiological data (ECG and EDA). Engagement annotations in this corpus are performed for each interaction session with a discrete Likert scale of $(1-7)$. 

\textbf{HCI.} Existing datasets in HCI are mostly recorded in the context of online learning. For instance, the DAiSEE dataset \cite{gupta2016daisee} includes learner's videos captured while they were watching a video tutorial, with a webcam mounted on a computer. It was collected in unconstrained conditions at different locations  with varying illumination settings.  Learner's engagement level  
as well as relevant emotions (bored, confused, and frustrated) annotations on a scale of $(0-3)$ are provided for each video. Similarly, ``in-the-wild'' dataset \cite{kaur2018prediction} for engagement assessment includes student's videos collected via Skype in an unconstrained environment. Engagement levels were annotated using crowd-sourcing in terms of 4 classes, including disengaged, barely, normally, and highly engaged. Finally, the HBCU dataset  \cite{whitehill2014faces} includes student's engagement level annotations assessed via crowdsourcing on data captured as the participants were engaged in a cognitive skill training study.

\textbf{HRI.} The Multimodal Human-Human-Robot Interactions (MHHRI) Dataset  \cite{celiktutan2017multimodal} was introduced for studying the relationship between personality and engagement simultaneously in dyadic HHI and triadic HRI. The context of the dataset is purely social revolving around personal questions asked by the interaction entities to each other. The dataset was recorded using biosensors, Kinect depth sensors in addition to first-person vision cameras attached to the participants heads. The engagement state of the users was assessed with a post-study questionnaire asking the participants about their perceived enjoyment of the interaction. In a later study \cite{salam2016fully}, labels from external annotators were obtained for this dataset using crowdsourcing. Another HRI dataset, the User Engagement in spontaneous HRI (UE-HRI)  dataset~\cite{ben2017ue}, was presented to study spontaneous social interactions between humans and a robot. The dataset includes 54 dyadic HRI interactions with the robot Pepper situated in a public space, collected over a period of $56$ days. The dataset was recorded using two 2D cameras, a 3D depth sensor, 4 directional microphones, sonar, and laser sensors. Recorded streams include face, speech, gesture, and dialog features. Engagement labels were obtained by external annotators.

In an educative HRI context, the Productive Engagement in HRI 
(PE-HRI) dataset \cite{nasir2020pe} is a multimodal dataset that allows studying  engagement in collaborative robot-mediated  educational contexts. The dataset includes productive engagement scores which are computed via a linear combination of the most discriminatory features \cite{nasir2022}. Additionally, the dataset consists of multimodal team level behaviours and learning outcomes (34 teams of two children). In a later version~\cite{nassir2021dataset}, the PE-HRI-temporal dataset was introduced where temporal features were computed in windows of 10 seconds for each team.

\subsubsection{Engagement Annotation}
\label{sec:annotation}
An essential step for building a reliable engagement inference system is acquiring the ground truth data. Engagement ground truth labels are usually assessed via validated or self-designed questionnaires such as the Temple Presence Inventory (TPI) \cite{lombard2011measuring} which was adapted and employed in HRI \& HAI contexts \cite{salam2016fully,campano2015eca}. 
Approaches to the collection of engagement ground truth labels can be divided in three categories: (1) \textit{self-report} labels, (2) \textit{external} measures, and (3) combination of self-report and external annotations.

\textbf{Self-report Labels.} These constitute pre- or/and during- or/and post-interaction self reports that gather information from the user about their experience with the technology. Using self-report evaluation of engagement can be considered as reliable since in theory one can truly know how they really felt during an experience. However, asking people to evaluate their experience after the experiment may be prone to error, as it relies on memory recall and on their attention to and communication of what made their sense of engagement to be perceived as powerful, or weak \cite{champion2003interaction}. 
Moreover, another issue that arises with post-experience questionnaires is that they do not take into account the interaction dynamics. Interactions are dynamic and change over time, making engagement to fluctuate~\cite{o2008user}. To account for these issues, some HRI works explored strategies to obtain the ground truth data from the users during the interaction by introducing implicit and explicit probes for collecting self-reported engagement ground truth labels at different stages of the interaction~\cite{corrigan2014mixing}. Also in the context of interaction with an online learning platform, emotional engagement self-labels were collected during the interaction via two modes: (1) voluntary where the students can provide their engagement label at any time via a window that appeared during the entire interaction on the interface and (2) mandatory via a pop-up window that appeared at random time intervals \cite{aslan2017students}.

\textbf{External Annotations.} These constitute recruiting a number of external annotators that assess an individual's level or state of engagement based on audio-visual recordings of the individual's interaction during the experience~\cite{nakano2010estimating}. External evaluation by external observers might be prone to errors due to the difficulty of perceiving the true engagement state of the user. In addition, it is often difficult to reach an agreement on the perceived engagement due to annotators' subjectivity. This is a general problem involving any social phenomenon, and there is already a line of work focusing on obtaining reliable ground truth labels from multiple annotations. For instance, in HRI,  independent  models  trained with different annotator's labels and then aggregated to obtain one integrated label~\cite{inoue2018engagement}. Similarly, the subjectivity of the annotators was considered by modeling each annotator's latent character affecting their engagement perception using hierarchical Bayesian model in~\cite{inoue2019latent}. The annotator's character and engagement level are estimated as latent variables. Experimental results show such modeling outperforms baseline models which do not take into consideration the differences in annotations and annotator's characters.

\textbf{Combining Self-report and External Labels.} As discussed above, both self-report and external annotations used to assess engagement ground truth suffer from certain disadvantages. To account for this, few approaches have attempted to combine self-report and external annotations to obtain engagement ground truth data. The hypothesis is that engagement prediction accuracy would be increased with a combined label reflecting a ground truth closer to reality, which is a combination between the perceived and reported engagement states. In the context of conversational HAI, past attempts for obtaining a combined self-report and external engagement label was to sum the user's and observer's judgments (e.g., user is labelled as disengaged when assessed as such by both parties) \cite{ishii2008estimating,nakano2010estimating,ishii2011combining}.

\subsubsection{Categorical View of Engagement} 
\label{sec:catview}
Most of the existing approaches to automatic engagement recognition formalise this problem as a classification task.  It is crucial to identify what classes are relevant to the interaction and application context. For instance, one might be interested in detecting the user's intention to engage  with one or all of the interaction parties, disengagement or the user's level of engagement. For this reason, we identify five categories of engagement in the state-of-the-art, summarised in Table~\ref{tab:eng_categ}.

\textbf{Intention To Engage.} Various engagement inference approaches concentrate on  detecting of user's intention to engage in the interaction \cite{xu2013designing,foster2013can}. Particularly, in HAI and HRI scenarios that include a social context, detecting the user's intention to initiate an interaction is of utmost importance for the agent to show an intelligent behaviour, and engage in a successful interaction. For instance,~\cite{foster2013can} focus on detecting user's intention to engage in an interaction with a robot using two classes: not seeking engagement vs. seeking engagement. 
Similarly, in \cite{xu2013designing} the occurrence of engagement intentions was modeled with two classes: E-intention (user wants to start a conversation with the agent or intends to speak, provided that the speaking floor was being held by others) and D-intention (user intends to disengage).

\textbf{Engagement/Disengagement.} Different approaches  focus on binary classification of engagement, aiming to detect the presence or absence of engagement vs. disengagement. For instance, in~\cite{leite2015comparing},  disengagement detection in individual and group interactions was investigated. On the other hand, linking engagement to attention, in~\cite{Yun2012} the focus was on the user's attention and lack of attention states.

\textbf{Engagement Process.} Several approaches focused on the classification of the phases of the engagement process, which constitutes mainly user's intention to engage, engagement, and disengagement. For example, in~\cite{benkaouar2012multi,vaufreydaz2015starting}, a model for recognizing 5 engagement classes was trained: no one, will interact, interact, leave interact, and someone around. 

\textbf{Engagement Level.} Another line of work deals with the recognition of user's engagement level or degree defining the engagement state on a spectrum, e.g. low to high. Examples of defined engagement level classes from the literature are summarized in Table \ref{tab:eng_categ} together with the respective context they where defined in.  

\textbf{Behavioural Engagement.} These approaches consider the behavioural aspect of the user during their engagement in an interaction. For instance,~\cite{Bednarik:2012} distinguish different states of conversational engagement: no interest, following, responding, conversing, influencing and managing. 
 Similarly,~\cite{salam2015multi,salam2015engagement} decomposed the engagement state into several mental, behavioural, and emotional states. In~\cite{oertel2013gaze} two classes in addition to high and low engagement levels were explored, namely, ``lead'' referring to when the game leader was directing the conversation and ``org'' corresponding to when the group was forming itself. In the context of team interaction~\cite{frank2016engagement}, the level of participation in a meeting was annotated with respect to six engagement states: disengagement (no participation, distraction and no attention to the meeting), relaxed engagement (attention to the meeting, listening, observing, but no participation), involved engagement (attention and non-verbal feedback), intention to act (preparation for active participation indicated through an increase in activity), action (speaking and/ or interacting with participants or content on displays), and involved action (intense gesture and voice).\\
\begin{table*}[htbp]
\scriptsize
  \centering
  \caption{Overview of the engagement categories that are widely used in the literature. IM: Interaction Mode}
  \label{tab:eng_categ}
\begin{tabular}{lllll}
\toprule
\textbf{Category} & \textbf{Paper}			& \textbf{Classes}& Context&\textbf{IM}\\
\toprule
\textbf{Intention To Engage} &\cite{foster2013can}&Not seeking engagement, Seeking engagement& &HRI\\
 &\cite{xu2013designing} & Engagement intentions: E-intention , D-intention , Attention saliency&&HAI\\
\midrule
\textbf{Engagement/Disengagement}& \cite{leite2015comparing}& Disengagement&&HRI\\
\midrule
\textbf{Engagement Process} & \cite{benkaouar2012multi}& Intention to engage, Engaged, Disengaged&&HRI\\
&\cite{michalowski2006spatial} &User is present, Interacting, Engaged, Just attending&Robotic Receptionist&HRI\\
 &\cite{vaufreydaz2015starting} & Will interact, Interact, Leave interact, No one, Someone around  & &HRI\\
\midrule
\textbf{Attention Level}&\cite{asteriadis2007non} &Distracted, Tired/Sleepy, Not paying attention, Attentive, Full of interest, Curious&Educative&HCI\\
\textbf{Attention/Frustration}&\cite{asteriadis2009feature}& Attention/Non attention, Frustration/Non Frustration&&HCI \\
\textbf{Attention}&\cite{Yun2012} &Attention, In-attention&&HRI\\				
\midrule
\textbf{Engagement Level}& \cite{castellano2012detecting}& Medium-high to high engagement, Medium-high to low engagement&Competitive&HRI\\
& \cite{oertel2013gaze}&  Group involvement:  High, Low, Lead, Group formation &Collaborative game&HHI\\
  & \cite{peters2010investigating}&   Engaged in the interaction, Superficially engaged with the scene and action space,&Agent salesperson&HAI\\
& & Uninterested in the scene or action space& \\
& \cite{corrigan2013identifying} &High and low engagement&Educative&HRI\\
\textbf{Interest Level}&\cite{kapoor2004probabilistic} &High interest, Low interest, Refreshing, Bored, neutral, Other&Competitive&HCI\\
\midrule
\textbf{Engagement Behaviour}
& \cite{Bednarik:2012}&  Conversational engagement:&&HHI\\
& & No interest, Following, Responding, Conversing,Influencing, Managing \\
& \cite{salam2015multi}& Intends to Engage,  Listening, Concentrating, Responding, Positive reaction, &Social, Informative&HRI\\ 
&&Negative reaction, Waiting feedback,  Thinking, Disengaged &Competitive&\\
\bottomrule
\end{tabular}
\end{table*}
 \vspace{-10pt} 
\subsection{Engagement Modalities and Features}
\label{subsec:feat}
\begin{figure}
    \centering
    \includegraphics[scale=0.5]{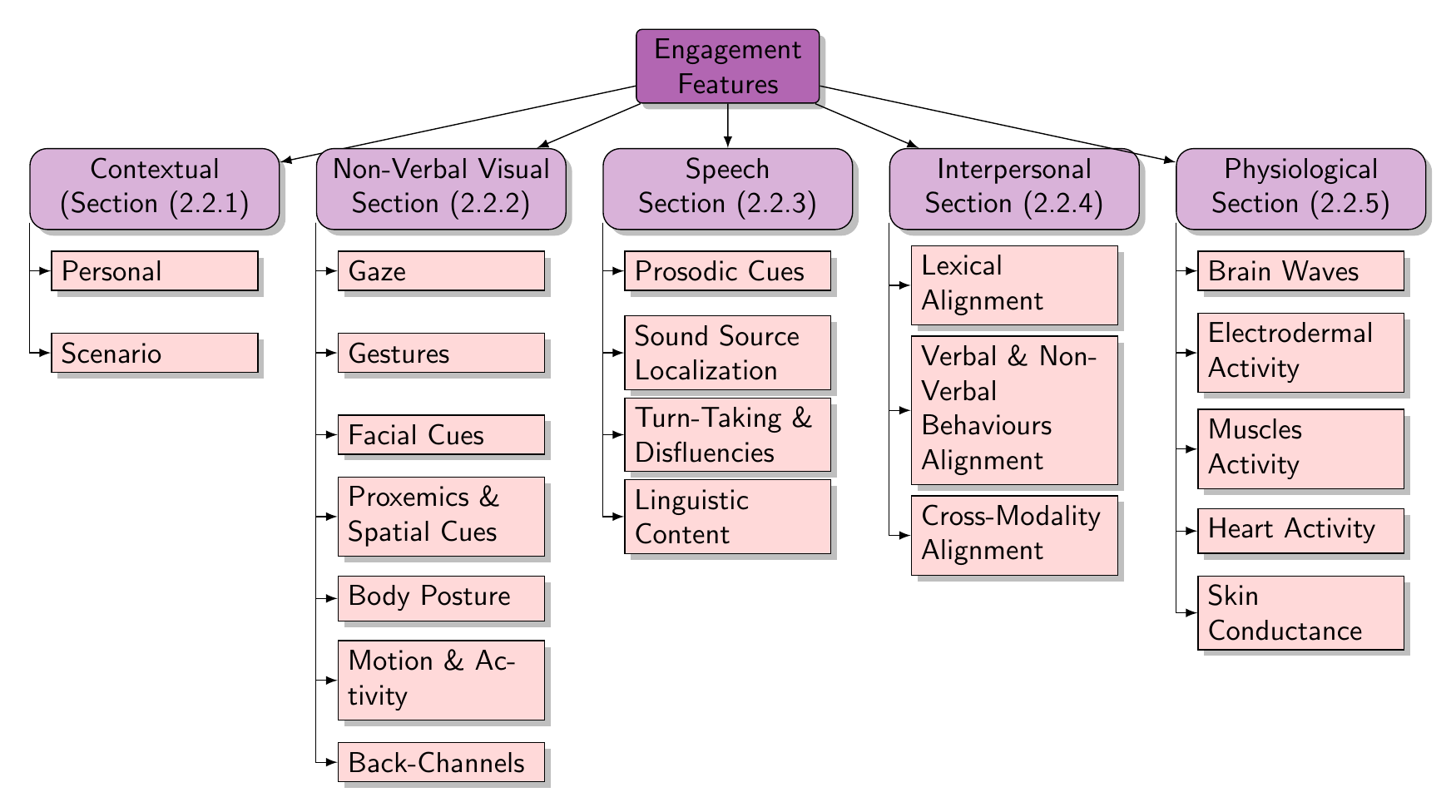}
    \caption{Engagement modalities and features categories. }
    \label{fig:featurestree}
\end{figure}

In the literature, a rich set of features were explored, extracted from various data modalities including audio, video, text, and physiological data, for engagement inference in different interaction contexts. While different categorisations can be found such as static vs. dynamic features~\cite{anzalone2015evaluating}, or low-level vs. high-level features~\cite{Glas2013,Atrey2010}, we divide the engagement features into five categories: (1) contextual cues, (2) non-verbal visual cues,  (3) speech cues, (4) interpersonal cues, and (5) physiological cues. These categories are depicted in a feature tree in section \ref{fig:featurestree}.

Our literature analysis of engagement modalities and features has shown that there is no observable trend of using specific features in specific contexts, as this depended on the available modalities and data, and engagement categories/scores,  which differed from one study to another. Nevertheless, in our review of contextual cues (Section \ref{sec:contextualcues}), we have found an important difference in the used contextual features among different interaction contexts (disembodied vs. embodied HMI). Consequently, in this section, we provide an overview of the features that are most commonly used in the literature of engagement inference. We include the context in which the different features are used, whenever it is relevant. We also discuss the difference in the used contextual features among different disembodied vs. embodied HMI interaction contexts.  
\subsubsection{Contextual cues}
\label{sec:contextualcues}
We have discussed in Section~\ref{sec:def} the importance of context for engagement. In this section, we review how context was used in the engagement inference literature, and what contextual features were extracted for this purpose across the different interaction modes (disembodied and embodied HMI) and scenarios. 

\textbf{Disembodied HMI.} Contextual features have been extracted depending on different {\bf interaction scenarios}. In an HCI \textit{competitive} game context, sequence mining was applied to combine physiological signals and game-context information for the prediction of computer game player affective states~\cite{martinez2011mining}. For detecting a user's attention level for tasks that commonly occur in a workplace setting, machine-specific contextual features were fused with person-specific features. The contextual features were collected from the keyboard, the mouse, and the active window size~\cite{sun2014nonintrusive}. In an \textit{educative} HCI scenario involving children engaged in solving a puzzle on a computer,  the game state was used as contextual features, and combined with facial and posture features for engagement inference\cite{kapoor2005multimodal}. In the context of interaction with a learning platform, contextual information were extracted from the URL logs and combined with appearance features to predict whether the user is engaged with the platform by analysing on-task vs. off-task behaviours \cite{okur2017behavioral}. 

\textbf{Embodied HMI.} In terms of {\bf interaction scenario}, in an HRI \textit{competitive} game scenario (robot vs. child), task and social-based contextual features in the form of robot's social behaviour, game events, and game progress were explored for the inference of engagement~\cite{castellano2012detecting} \cite{castellano2017detecting}. In a multiparty HRI scenario with varying contexts including \textit{informative}, \textit{competitive}, and \textit{social}, \cite{salam2015engagement}, the robot's and other participant's behavioural cues were used as contextual information to detect the user's engagement state. The study has shown that behavioural cues from the other parties (robot and other participant) can be used as predictors of the engagement state of the user in question. This might happen when in multi-party interactions, there is a significant cohesion and synchrony within the group, which might signal high engagement.  Similarly, in \cite{baur2016modeling}, to integrate context into the proposed framework of engagement inference, the current user's engagement is modeled as a function of previous user's engagement state. The approach is validated in the context of \textit{negotiation} scenario (i.e., multi-agent job interview) and a \textit{social} robot scenario. 

In terms of {\bf personal factors}, a line of work investigated the impact of personal cues such as individual's \textit{personality} as well as their interaction partner's personality on the inference of engagement~\cite{salam2016fully} or rapport~\cite{7439777}. Salam~\etal~\cite{salam2016fully} focused on a triadic HRI interaction scenario (i.e., between two humans and a robot) and showed that using individual's personality traits only was sufficiently informative to detect their engagement, and combining those with interpersonal features resulted in further improvement in the accuracy of engagement inference. Cerekovic~\etal~\cite{7439777} analysed the impact of personality on the perception of rapport in an HAI scenario. Participants interacted with two different agents, displaying different characteristics (i.e., cheerful vs. gloomy), and they found that extroverted and agreeable people tend to report higher levels of rapport with both agents. 
 \textit{Culture} and \textit{gender} were used implicitly for  personalizing engagement recognition deep architectures to specific population of differing gender or culture \cite{rudovic2018culturenet}, \cite{rudovic2019multi}, \cite{rudovic2019personalized}. We touch into this more in detail in the Section \ref{sec:deeplearning}, reviewing personalized deep learning models for engagement inference.

Despite evidence on the predictive power of contextual information for engagement inference, however,   methods using contextual features remain scarce in the literature.  
Devillers \etal \cite{devillers2017toward} highlight the importance of taking into account context in the assessment of engagement in HRI, namely the behaviour of the robot, the interaction dynamics, and the dialogue participants communicative behaviour variations. They identified linguistic, paralinguistic, interactional, non-verbal, and specific emotional and mental states features to be of utmost importance for engagement prediction.

\subsubsection{Non-verbal Visual Cues}
Non-verbal cues comprise social signals exchanged during social interactions along with the uttered verbal words. This type of behaviour is the most informative, when individuals can easily verbally pretend that they are engaged in a task or in a conversation. However,  it is much harder to fake non-verbal behaviour which is indicative of engagement. 
Some researcher even claim that initiating and maintaining  engagement is fully possible without verbal conversation~\cite{Sidner2002}. Features in this category include visual cues (eye gaze, gestures, facial expressions, proxemics and spatial features, body posture, and motion and activity features), and audio cues (prosodic, sound source localization, turn-taking, and back-channels). Commonly used non-verbal visual cues include: (1) gaze, (2) gestures, (3) facial cues, (4) proxemics and spatial cues, (5) body Posture, (5) motion and activity, and (6) non-verbal back-channels.  

\textbf{Gaze.} Eye gaze is the mirror to the mind and is strongly coupled with cognitive and emotional processes. It is considered as the primary cue of attention \cite{mason2005look}, a cognitive component of engagement. For instance, engagement with a speaking conversational partner can be signaled by gaze cues such as looking at them. On the other hand, looking away from them for long periods indicates disinterest and consequently a disengagement intention \cite{sidner2003engagementwhenlooking}. 
Gaze is among the most commonly used features for inferring engagement across various interaction contexts. Employed gaze features include gaze direction \cite{Yun2012,ehrlich2014engage,Papadopoulos2013,castellano2012detecting}, visual focus of attention \cite{Klotz2011,salam2016fully}, gaze transition patterns~\cite{nakano2010estimating,ishii2008estimating,ishii2011combining}, the amount of time the user's gaze was directed at the interaction partner \cite{Castellano2009,sidner2005explorations}, gaze following of the subject of speech \cite{sharma2014me}, and mutual gaze \cite{rich2010recognizing}, and gaze fixation duration \cite{bekele2014multimodal}. 
Eye gaze and head orientation are closely linked and often one accompany the other. Pointing gestures can also contribute to the computation of the other's direction of attention~\cite{Langton2000eyes}.
In some scenarios, it might be difficult to detect the eye gaze accurately (e.g., due to low resolution images, adversarial head poses, or occlusions). Several researchers, opted to approximate the user's gaze by the head pose \cite{benkaouar2012multi,sidner2005explorations,Sidner2006,Bohus2009,foster2013can,Papadopoulos2013}, especially in HRI where there is a big distance between the robot and the interactant.

\textbf{Gestures.} 
Gestural behaviour in interactions conveys a lot information on the engagement state and on the connectedness between participants~\cite{mcneill1992hand}. 
Moreover, in HAI and HRI the generation of appropriate gestures via the agent/robot and the successful interpretation of human gestures have a significant effect on the user's engagement and consequently the interaction success~\cite{sidner2003architecture}. Relevant gestures include head gestures such as head nods and shakes (convey (dis)agreement and attention) \cite{kapoor2005multimodal,Sidner2006}, and hand gestures and speed such as  hand raising (convey intention to engage) \cite{tofighi2016vision,frank2016engagement,ts2020automatic}. 

\textbf{Facial Cues.} As discussed in previous sections, engagement has an affective component. Consequently, facial expressions and emotions are important features for detecting user's engagement state across various interaction modes and scenarios. Some approaches focused on a small class of expressions, such as the smile \cite{castellano2012detecting,frank2016engagement} or the  probability of smile and fidget \cite{kapoor2005multimodal}. 
Other approaches detected multiple emotion states \cite{thiruthuvanathan2021engagement} and facial expressions such as smiling, surprised, neutral, and angry states \cite{Yun2012}. Fine-grained facial movements such as the displacement and velocity of the mouth and eye landmarks \cite{De2004}, or facial action units (AUs) were also used as engagement features \cite{vail2014predicting,whitehill2014faces,Grafsgaard2014}. \cite{Monkaresi2014} concluded that eye's region is more informative than mouth's region. Eye physiological indices such as pupil diameter and blink rate and eye behavioural indices are also relevant to engagement inference \cite{bekele2014multimodal}.

\textbf{Proxemics and Spatial Cues.} Spatial behaviour (proxemics) constitutes the dynamic process by which individuals position themselves in social interactions~\cite{hall1959silent}. Proxemics carry significant attentional and interpersonal factors \cite{mead2011proxemic}, which are indicative of an individual's engagement state. For instance, being at a relatively small distance indicates higher probability of paying attention to a robot \cite{michalowski2006spatial}.  
 Holthaus \etal \cite{Holthaus2011} proposed a spatial model for a receptionist robot to infer the user's intention. Other features to consider are the relative distance between the interaction partners \cite{salam2016fully}, or between the user and the robot or the machine \cite{Yun2012,vail2014predicting,Bohus09}, as well as the trajectory or speed of a person such as walking towards a robot. The size of the face detected by the camera can also be used to approximate the distance between a person and a robot \cite{benkaouar2012multi,foster2013can,Papadopoulos2013}. Detecting  if the face is currently invisible, and for how long was also used to  indicate a person's spatial situation \cite{bohus2014managing}.

\textbf{Body Posture.} 
Previous research has shown that the face's and body's orientation towards an interlocutor is a signal of engagement~\cite{Peters2005,fast1970body}. The body orientation and relaxation were also found to be significant indicators of the communicator's liking for their addressee~\cite{mehrabian1968}. The dynamics of postural shifts might even signal shifts or changes in the conversation topic (boundaries of information units) during an interaction~\cite{cassell2001non}. Moreover, as the posture is often displayed unintentionally, this makes it an effective indicator of the user's real affective state~\cite{person2008toward,kleinsmith2013affective,witchel2014time}, as well as their attitude and alertness \cite{bull1983body}. 
Various approaches employed body posture features in the literature of engagement inference \cite{kapoor2004probabilistic}. Among such features, the upper body orientation, the back's curvature, and the upper body degree of contraction and expansion (contraction index) were used in the context of human-child game interaction \cite{sanghvi2011automatic}. Naturally occurring seated postures were also used as features for child engagement during a computer learning task \cite{mota2003automated}. Body lean angle, slouch factor, body direction, hand vertical position, and posture were also explored to detect engagement in team meetings~\cite{frank2016engagement} and student engagement with MOOCs \cite{kaur2019domain}. Other features include hand pose and body orientation \cite{foster2013can}, upper body posture and body openness \cite{baur2013nova}, the position and orientation of the feet,  hips, and torso and the shoulders positions and orientation relative to each other \cite{benkaouar2012multi}, the relative orientation between the participants and the robot \cite{salam2016fully}.

\textbf{Motion and Activity.}
Previous studies found correlations between the user's quantity of movement and their degree of involvement in an interaction \cite{Oertel2011,witchel2014does}. It is actually possible to judge users' level of engagement by measuring their movements as they use a computer \cite{witchel2014does}. For instance, increased user movement can be an indicator of an increased involvement, consequently an increased engagement state \cite{Oertel2011}. On the other hand, the absence of certain  movements can signal cognitive engagement  in seated situations \cite{witchel2014does}. In \cite{witchel2014does}, the authors   distinguished between instrumental movements referring to physical movements serving a direct goal in a given situation, and non-instrumental movements that are involuntary tiny movements that people usually exhibit to reflect the person's internal states. If someone is absorbed in what they are watching or doing, referred to as ``rapt engagement'' by \cite{witchel2014does}, there is a decrease in these non-instrumental movements.
Motion features used for engagement inference include the Quantity of Motion (QoM), which is a measure of the amount of detected motion. QoM was used by \cite{sanghvi2011automatic,salam2016fully} to infer engagement with a robot companion. 
 The approach of~\cite{salam2016fully} also detected the global QoM of a multi-party interaction for group and individual engagement inference. The body activity computed from skeleton joints of upper body bounding boxes was also used in this context~\cite{salam2016fully}.

\textbf{Back-Channels}. 
Back-channels correspond to events where an interaction party responds back to a primary communication initiator with a brief verbal or gestural communication \cite{rich2010recognizing}. Example non-verbal back-channels include head nods and shakes signaling to the initiator that the responder understands, listens or desires to continue the conversation. In HHI, head nods were combined with other non-verbal features to detect team engagement in meetings~\cite{frank2016engagement}. In HRI, ~\cite{sidner2005explorations} a robot was endowed with the capability of interpreting nods as back-channels and agreements in conversation in order to recognize the user's state of engagement. Back-channels together with laughing and nodding were also found to be related to the level of engagement in social HRI scenario \cite{salam2015engagement}.

\subsubsection{\textbf{Speech Cues}}
Speech cues  are used in the context of HMI for engagement recognition, since some spoken linguistic behaviour might convey  engagement~\cite{sidner2005explorations}. Speech cues can be extracted from verbal speech, or written text. Other than the linguistic content, various non-verbal signals are encoded in speech. The analysis of written or spoken language is interesting for the context in question because it permits a straightforward way to give an input to the engagement system: greetings can be considered as a cue for intention to engage~\cite{sidner2003engagementwhenlooking,Bohus2009}, while closing comments may signify the disengagement of the user~\cite{sidner2005explorations}. Using verbal behaviour, the user has an active role during the interaction in a way that would not be possible if only non-verbal visual behaviours were considered. Embodied HMI (HRI and HAI), for instance, can be more natural if humans can speak directly with robots/agents. Several works consider acoustic and text-based data to obtain information about the user's engagement \cite{Maisonnasse2007}.  These can be categorized into: (1) prosodic cues, (2) sound source localization, (3) turn-taking and disfluencies (4) and linguistic content. 

\textbf{Prosodic Cues.} Previous research~\cite{Oertel2011} observed a relationship between the voice's prosodic characteristics (level, span, intensity) and  involvement. The authors of~\cite{yu2004detecting}, for example, used sound and prosodic features such as the speech rhythm, stress, and intonation, to estimate conversational engagement level between  a voice communication system users. Another work~\cite{frank2016engagement} used the speech volume in combination with other non-verbal features to predict team engagement in meetings.

\textbf{Sound Source Localization (SSL).} Detecting the sound source can be an indicator of user's intention to engage, engagement or disengagement. For example, the robot's ability of localising someone next to it from the sound, identifying speech activity or even prosody allows it to recognise the engagement state of a user. SSL was used in the literature for locating the user's voice or footsteps, allowing to detect the direction in which the user approaches a robot, which indicates their intention to engage \cite{benkaouar2012multi}. SSL in an HRI setting~\cite{kim2004reliable} was also used to estimate user's intention to engage and engagement level in~\cite{Maisonnasse2007}.

\textbf{Turn-Taking.} The notion of turn-taking also plays an important role in the context of conversational HAI and HRI. As Sidner~\etal~\cite{sidner2003engagementwhenlooking} observed, failure of an interaction party to take an expected turn is an indication of disengagement. Thus, related to this notion, adjacency pairs which consist of ``two utterances by two speakers, with minimal overlap or gap between them, such that the first utterance provokes the second utterance'' were used  as an engagement cue in \cite{rich2010recognizing}. 
 Similarly, engagement annotation of conversational data was studied in the context of HRI, where it has been demonstrated  that engagement level is correlated with turn-taking behaviours such as the duration of the next turn \cite{inoue2016annotation}. Turn-taking features employed in the literature include speech and pause duration statistics, speaker change with and without overlap, successful and unsuccessful interruption, and speech overlap \cite{kim2016multimodal}.

Disfluencies, such as filled pauses (also referred to as fillers), short and long speech pauses between words, or hesitations were also considered in the literature of engagement detection and management. For instance, combined with gaze, filled pauses (e.g. ``uh'' or ``umm'') were  used to detect whether a user wishes to disengage from the interaction with a robot \cite{bilac2017gaze}. In \cite{bohus2014managing}, linguistic hesitations (e.g. filled and non-filled pauses) were used for managing conversational disengagement in HRI when uncertainty about whether the user wishes to stay engaged in the interaction arises. In a robot-mediated collaborative learning context involving teams of two children, speech behaviors including the team members speech overlap or the amount of their long pauses were found to be the most discriminating between teams exhibiting high learning engagement and teams exhibiting low learning engagement. An  engagement score was  generated using a linear combination of such behaviours \cite{nasir2022}.

\textbf{Linguistic Content}. Certain spoken words can signal  engagement and related affective states~\cite{Strapparava2004}.
The use of linguistic content in engagement inference is  scarce in the literature. Example works include~\cite{foster2013can} who presented a method for speech recognition using grammar implementation. The proposed method aimed to extract syntactic and semantic information from the user's speech to detect engagement. In~\cite{vail2014predicting}  a tutorial scenario was proposed to predict engagement and learning, where the user was able to send textual messages to a virtual agent. 

\subsubsection{Interpersonal cues}
Interpersonal features represent the interpersonal behaviours of the interaction parties relative to each other. The cues discussed in the previous sections can be considered as individual cues that are extracted by isolating the interaction partners or action-reaction processes where interaction partners exhibit a behaviour in response to each other (e.g., turn-taking, back-channels). In this section, we look at more elaborate cues such as synchrony or alignment. Synchrony is defined by \cite{delaherche2012interpersonal} as ``the dynamic and reciprocal adaptation of the temporal structure of behaviours between interacting partners''. Unlike mirroring or mimicry, synchrony is dynamic in the sense that the important element is the timing, rather than the nature of the behaviours \cite{delaherche2012interpersonal}. Several studies in HMI have referred to the importance of synchrony for engagement inference. Detecting synchrony between the interaction parties involves not only the detection of their reactivity, but also their agency and their engagement within the ongoing interaction~\cite{nadel2005experiencing}. For instance, Prepin and Gaussier \cite{prepin2010agent} showed that synchrony is a viable indicator of the user's satisfaction and level of engagement during an interaction with a robot. The better the interaction is, the more the human is synchronous with the robot. They designed a robotic architecture that can detect temporal synchrony between the user and agent's actions and use this parameter to adapt the robot's behaviour using reinforcement learning. 

Three categories of approaches that use participant's synchrony for engagement recognition can be identified in the literature: (1) lexical, (2) verbal and non-verbal behaviour, and (3) cross-modality alignment. 

\textbf{Lexical alignment} refers to the adoption of one’s interlocutor’s lexical items~\cite{foltz2015lexical}. State-of-the-art approaches investigated lexical alignment by detecting the use of shared vocabulary and verbal repetitions~\cite{campano2014comparative} as a cue of engagement in an attempt to develop a virtual agent capable of employing alignment strategies to maintain user's engagement. 

\textbf{Verbal and non-verbal behaviour alignment} refers to the alignment or synchrony between participants verbal or non-verbal cues during a social or collaborative interaction. Studies in HHI found similar inphase/antiphase dynamic among interaction parties engageing in an interpersonal task \cite{Richardson2005Towards}. In the context of multi-party HRI, Salam~\etal~\cite{salam2015engagement} showed that it is possible to detect the engagement of an interaction party using as input the behavioural cues of the other interaction parties (robot and other participant) with an acceptable accuracy compared to solely using the interaction party individual behavioural cues. Such approach can be considered an indirect way of using synchrony features since the high correlation of others' cues with the engagement of the participant in question means that they were in synchrony with her. They~\cite{salam2015engagement} also extracted a set of features describing the synchrony and alignment between robot's and participants' behaviours. These include, among others, mutual gaze and laughter. 
Other features included events where a participant speaks with the other during the speech of the robot. This may signal a disengagement behaviour as it means that the participant is not listening to what the robot is saying.

\textbf{Cross-modality alignment} refers to the alignment or synchrony between different modality cues such as speech and gaze, or speech and gesture. Findings suggest that gazing towards objects relevant to the conversation is an indicator of engagement~\cite{sidner2003engagementwhenlooking}. For instance, in the context of a multi-party HRI scenario~\cite{salam2015engagement}, events where one participant looks at the other who is speaking to the robot or events where a participant looks at objects corresponding to the topic of robot's speech (i.e., gaze-speech alignment) were used as predictors of the user's engagement state. 
In \cite{baur2016modeling}, a study was performed in the context of two scenarios:  a social robot scenario and a multi-agent job interview scenario,  
proposing to model the   interpersonal cues dynamics that reflect the social attitude of the user with the context. The user’s engagement was considered as a combination of the user’s individual behaviour patterns and their interpersonal behaviour patterns as well as their temporal alignment. 

\subsubsection{Physiological cues}

Previous research provided evidence on the relationship of physiological signals with affective and cognitive states.  Composed of both affective and cognitive components, engagement can be successfully predicted from physiological responses.
Various state-of-the-art methods have employed physiological signal analysis for the inference or analysis of engagement or any of its related constructs. 
 For instance, in~\cite{jang2015analysis}  physiological features were extracted from Electrocardiogram (ECG),  electrodermal activity (EDA), and Photoplethysmography (PPG) signals to differentiate between boredom (associated with disengagement), pain and surprise. 
 In~\cite{ehrlich2014engage}, EEG signals were used with the goal to capture brain activity for understanding the link between intention and attention. In~\cite{Belle2012}, EEG, ECG, and Heat Flux (HF) signals were investigated for the identification and evaluation of engagement level variations during cognitive tasks. 
 In the context of tele-operated HRI,  human engagement (attention) on a specific task was also detected from physiological signals~\cite{rani2007operator}. The extracted features included cardiac activity, heart sound, Bioimpedance, EDA, muscle activity and  Skin Temperature (SKT). Results showed that the correlations between engagement ground truth labels and features were not similar among all participants indicating that people do not express engagement in the same manner.
 In~\cite{Monkaresi2014}, video-based heart rate (HR) measurement combined with appearance and geometric features was explored for engagement inference. In \cite{fairclough2005psychophysiological}, it was found that a notable percentage of the variance for task engagement can be predicted from ECG, electro-oculogram (EOG), skin conductance (SC), and respiratory rate (RR) signals. In~\cite{iani2004effects}, it was shown that blood volume pulse (BPV) amplitude varies significantly among task and resting periods, which signals variations in sympathetic activity when engaging with a task.

The advantage of using physiological signals for predicting the affective and cognitive components of engagement is that they are able to provide precise measurements, since they give an objective insight on the true state experienced by the user. However, their intrusive nature requiring to attach various sensors to the user's body limits their use in natural human-machine interaction settings and their scalability to a wide range of applications. Another important issue  with physiological signals comes from the inter-intra individual variability.

\subsection{Engagement State Inference}
\label{subsec:fus}
Engagement state inference methods can be classified into three different categories: (1) rule-based, (2) supervised learning and (3) unsupervised learning. The employment of  traditional machine learning techniques was more prevalent prior to the emergence of deep learning methods and their success in various pattern recognition problems. On the other hand, unsupervised learning approaches to engagement inference were also explored in the literature. However, these are limited to educative HCI contexts.

\subsubsection{Rule-based Approaches}

Rule-based approaches for engagement inference infer a decision on the user's engagement state based on a set of custom-defined  rules (IF some condition THEN some action). Most of the rule-based engagement inference methods can be found in HRI contexts. This is mainly because rule-based methods are less expensive in terms of resources requirements, making them more suitable for integration is real-time settings. Sidner \etal \cite{sidner2003architecture} implemented one of the first robotic architectures that is endowed with engagement rules to generate appropriate engagement behaviour for the robot, and to detect the human's engagement state enabling a successful  collaborative conversation between the human and the robot.
 Similarly \cite{sidner2005explorations} endowed their robot with rule-based engagement inference capabilities. The robot judges the user's engagement based on the head's position which indicates if the user is looking at the robot, at objects necessary for the collaboration or to other objects or to empty space.
Similarly, \cite{maniscalco2022bidirectional} implemented a multimodal rule-based real-time robotic architecture that detects user engagement based on movement detection (head and body tracking), face recognition, gaze direction,  proxemic distance, and audio cues (sound direction localization, audio signal power). Other works that employed rule-based engagement inference methods include  \cite{rich2010recognizing},  \cite{Klotz2011}, \cite{foster2013can}, \cite{Bohus09,bohus2014managing}. Comparing rule-based user  engagement classification to machine learning methods, \cite{foster2013can} found that trained classifiers were faster and more accurate at detecting the user's intention to engage,  while the rule-based approach resulted in more stability. 

An important advantage of rule-based engagement inference approaches is explainability, an important aspect to take into account to avoid potential unwanted bias towards certain protected social groups (e.g. gender or race). Other advantages include  the simplicity and rapidity of implementation, and the non-necessity of large training datasets. On the other hand, such advantages come at the cost of ignoring important complex patterns that can be automatically discovered from data by machine learning models, compromising the accuracy of rule-based methods.

\subsubsection{Supervised Learning Approaches}
Supervised learning methods were widely used for engagement inference in the literature. We categorize the employed supervised learning methods into two categories: (1) traditional machine learning, and (2) deep learning approaches. 

\paragraph{\textit{Traditional Machine Learning Approaches}}
A range of classification techniques  have been used to classify multi-modal observations into one of the pre-defined engagement classes. The performance depends on the general framework and used modalities.
Examples include Support Vector Machine (SVM) \cite{castellano2012detecting}\cite{vail2014predicting}\cite{ehrlich2014engage}\cite{benkaouar2012multi}\cite{whitehill2014faces} \cite{Monkaresi2014}, GentleBoost, AdaBoost  \cite{whitehill2014faces}, Multinomial logistic regression \cite{bohus2014managing} \cite{whitehill2014faces}, General Regression Models (GRM), C4.5 (decision tree) \cite{Monkaresi2014}, and Random forests \cite{Belle2012}, Dtree and OneR  \cite{sanghvi2011automatic},  boosted decision trees models \cite{bohus2014managing},   Fuzzy min-max neural networks (FMMNN)  \cite{Yun2012}.  
Comparing various classifiers, \cite{foster2013can} concluded that trained classifiers are faster and more efficient compared to rule-based methods. However rule-based methods are more stable (prediction variables vary less frequently during an interaction). Moreover, in order to obtain better results, it's possible to add some temporal features to the states using Conditional Random Field or HMM. For instance, employing a multilevel structure with coupled HMM led to suitable performance for engagement inference in a usual daily conversations \cite{yu2004detecting}. In \cite{mota2003automated},   learner's engagement level was estimated using a combination of neural networks  and Hidden Markov Models.
Other probabilistic methods that were employed in the literature include Bayesian inference methods \cite{Castellano2009},  \cite{baur2013nova} and Sugeno-type fuzzy inference system \cite{asteriadis2009feature}.

\paragraph{\textit{Deep Learning Approaches}}
\label{sec:deeplearning}
The rise of deep learning and its track record in achieving high performance for various affective computing problematics has led researchers to propose various deep learning approaches for engagement recognition. A clear added value of deep learning is their ability to learn new featureres presentations. This is also a nice way to learn mixed representations using contextual information, which is harder with other approaches.
The most recent deep learning approaches  concentrate on students engagement inference in online learning. This is due to the recent shift to online learning due to the Covid-19 pandemic. 
Deep learning approaches for engagement inference  either use raw data, or  behavioural cues as input to deep architectures.

In HRI,  various architectures composed of  Convolutional Neural Networks (CNNs) and  LSTM were proposed for engagement classification or regression. For instance \cite{hadfield2019deep} proposed to classify engagement using body pose estimated from multiple RGB depth cameras as input to an LSTM layer.
Other architectures composed of  CNNs followed by LSTM \cite{del2020you} were also employed to predict a continuous engagement score  from robot-view video streams.

In online learning context, an approach for student engagement prediction has been proposed by fusing facial and body features into a single long short-term memory (LSTM) model \cite{li2019feature}.
Dilated Temporal Convolutional Networks (TCN) has also been proposed for predicting student engagement intensity\cite{thomas2018predicting}. It has been demonstrated that TCN capture long term dependencies and it outperforms other sequential models like LSTMs. 

A recent trend, is the development of personalized engagement models. 
ResNet-50 architecture  \cite{rudovic2018culturenet} was   proposed to train culture-wise personalized engagement models (CultureNet) from face images for engagement inference in the context of robot-assisted therapy for autistic children.  \cite{rudovic2018personalized} proposed a personalized multitask learning framework to simultneously predict engagement, valence and arousal. The network learns behavioural multimodal (visual, audio, physiological) features representations using autoencoders. Personalization with respect to culture, gender and each individual is performed using different fine-tuning strategies.

\subsubsection{Unsupervised Learning Approaches}
There has been an interest in developing unsupervised learning approaches to engagement inference, although mainly these approaches are based clustering techniques and are limited to HCI and education context. These approaches aim to discover learners' engagement patterns including engagement state, level or style from data, e.g., system log files.

\textbf{Engagement State}. A line of work has focused on identifying patterns corresponding to a binary engagement state of the learner, namely, engagement vs. disengagement. For instance, in the context of MOOC, Coffrin~\etal~ \cite{coffrin2014visualizing} were able to differentiate between engagement and disengagement in the course based on student's learning analytics including histogram of student's performance and weekly student's participation. 

\textbf{Engagement Level}. Some approaches looked into detecting the level of user's engagement with the learning environment. For instance, in the context of exploratory learning environments \cite{amershi2009combining}  unsupervised clustering (k-means) were applied to identify interaction patterns corresponding to different levels of learner engagement from gaze and context-based features  reflecting students actions during their learning experience. Two engagement levels were identified, namely, high and low learners.  

\textbf{Engagement Style}. These approaches aim to identify learners engagement style or behaviour based on interaction patterns with the learning environment.
 For example, in \cite{kizilcec2013deconstructing,anderson2014engaging}, learners were clustered based on their degree of lectures and assignment completion. 
In the context of robot-mediated learning \cite{nasir2022many}, approaches were designed using forward and backward clustering from the multimodal behavioural features to the learners learning outcome metrics and vice-and-versa, allowing the identification of learner profiles (gainers vs. non-gainers). An engagement score was then generated using a linear combination of the most discriminating  behaviors  between the identified gainers and non-gainers profiles \cite{nasir2022}. 

\section{Discussion and Open Questions}

In this section we discuss the main points presented in this survey paper as well as  open questions in automatic engagement perception and modeling which remain under-explored and deserve attention.

\subsection{Context-aware Engagement Modeling}
One important question, related to context-aware engagement modeling,  is how to incorporate contextual information in the automatic engagement inference systems. 
For instance, in HRI, the work of \cite{devillers2017toward} highlighted the  importance of accounting for high-level and low-level  communication processes when measuring engagement. In addition to  communicative behaviour cues (e.g., visual, linguistic), and interpersonal cues (communicative behaviour of interaction parties w.r.t each other), it is necessary  to consider the interaction dynamic (i.e. variations in interaction parties communicative behaviour ), as well as  contexts (e.g., human-robot relationship, social, situational, human profile).

As discussed in Section \ref{sec:contextualcues}, most of the existing approaches incorporated contextual information in the form of features extracted from the task or from the interaction parties' behaviours. However, research has shown that the context of the interaction may evolve over time,  impacting the interaction parties' behaviours and consequently, the engagement models' accuracies  \cite{salam2015multi}.  
Regardless of the ultimate interaction goal, various interpersonal sub-contexts (e.g. social, informative, etc.) \cite{salam2015multi} might emerge during the same interaction, which might trigger different cognitive, emotional and behavioural user states, indicative of their engagement state. While previous studies relate some  mental states to engagement, the literature lacks a clear indication of when a certain state (e.g emotional, cognitive) is a significant indicator of engagement in a certain context. It is thus necessary to understand how the engagement cues emerge and fluctuate during the same interaction in relation to the sub-context, and to consider such variations when developing automatic engagement inference models. 


\subsection{Temporal Dynamics of Engagement}
\label{sec:temporaldynamics}
Previous works have indicated that engagement is a dynamic process that changes over time and that is dependent on the participants of a continuously evolving interaction~\cite{glas2015definitions,devillers2017toward}. This dynamic aspect was not thoroughly studied in the literature of automatic engagement inference. As a matter of fact, we find different ways by which researchers segment their videos and address the problem of timescale without giving any justification on why the specific time scale was chosen. For instance, \cite{vaufreydaz2015starting} detected engagement in 80 milliseconds video segments, \cite{xu2013designing} used 0.5 second fragments, and  \cite{kapoor2004probabilistic} segment their videos into a maximum of 8 seconds fragments.
One study \cite{whitehill2014faces}, has addressed this matter in the context of student engagement and performed a comparison between 1 frame, 10 second segments  and 1 minute segments engagement labeling by external annotators. They found that the labeling task was easier and more reliable (high inter-rater agreement) when annotators labeled 10 second video clips, and that a reliable prediction of  engagement labels of 10 second video segments can be obtained from the average of their constituting frames labels.
Consequently, an important aspect to take into consideration in engagement inference is the optimal observation window in which engagement can be detected. Relevant research questions that are still open  for investigation include the following. Is it sufficient to perform a \textit{static} (frame-level) inference? If yes, is this achieved by re-using clip-level engagement label as the labels for the constituent frames? What are the advantages and disadvantages of this? Or a \textit{dynamic} (segment-level) inference is more relevant? 
In the case of dynamic inference, what is the optimal time window, and is this  context-dependent? Is engagement in specific time segments or frames affected by the users' past behaviors, and if yes, to what extent? Such questions merit further investigation in future studies. 
\subsection{Personalised Models and Bias}
As discussed in Section \ref{sec:deeplearning}, a recent trend in engagement prediction systems is the development of personalized models. In such frameworks, user profiling  w.r.t personal factors (cf. Section \ref{sec:personalfactors}) can be performed prior to training the models. Such information can then be used at the level of the data, or within the machine learning process to adapt the models to the specific profiles. Compared to one-fits-all approaches which are simpler to train but can compromise the models accuracy and adaptability, personalized models seem to be promising. However, they remain under-explored. 
One concern that arises in this context is the problem of bias and potential unfairness of personalized models decisions to certain social groups (e.g. gender, age). Generic models of affect were shown to present certain biases to such groups \cite{XuWKG-ECCVW20}, if not properly tackled \cite{cheong2021hitchhiker}. However, bias investigation and mitigation in generic engagement models is still not explored in the literature.
Moreover, the question of whether personalization of automatic engagement inference systems increase or mitigate bias and fairness remains open, and merits further investigation.

\section{Conclusion}
\label{sec:conclusion}
In this paper, we presented a context-driven survey on engagement in human-machine interaction across various modes of interaction (HHI, HCI, HRI, and HAI). We reviewed more than 200 papers and we introduced widely used engagement definitions, available datasets, widely used features, and machine learning approaches. Engagement is a key component of social intelligence. We believe our survey will be a helpful guide for researchers working or planning to work on the problem of engagement inference, and aiming to equip machines with social intelligence. We finalised our survey with discussions and open questions to present our insights into how to advance this area of research further.

\ifCLASSOPTIONcaptionsoff
  \newpage
\fi

\bibliographystyle{IEEEtran}
\bibliography{bibtex/bib/TAC_SurveyEngagement}

\begin{IEEEbiography}[{\includegraphics[width=1in,height=1.25in,clip,keepaspectratio]{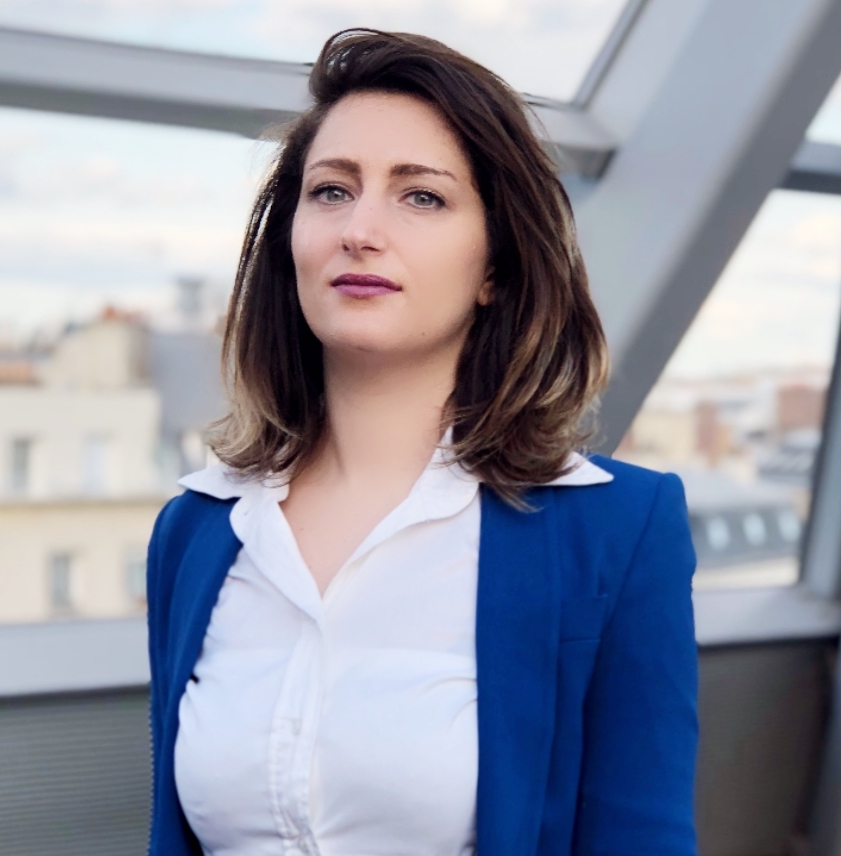}}]{Hanan Salam} is Assistant Professor in Computer Science at New York University Abu Dhabi. She is also the director of Social Machines \& Robotics Lab (SMART) \& a member of the Center of AI \& Robotics (CAIR). She is the co-founder of Women in AI, an international non-profit whose mission is to close the gender gap in the domain of Artificial Intelligence through education, research, and events. Her scientific interests include Artificial Intelligence for mental healthcare, Human-Machine Interaction, social robotics, computer vision, machine learning and affective computing. 
\end{IEEEbiography}

\vskip -2\baselineskip plus -1fil
\begin{IEEEbiography}[{\includegraphics[width=1in,height=1.25in,clip,keepaspectratio]{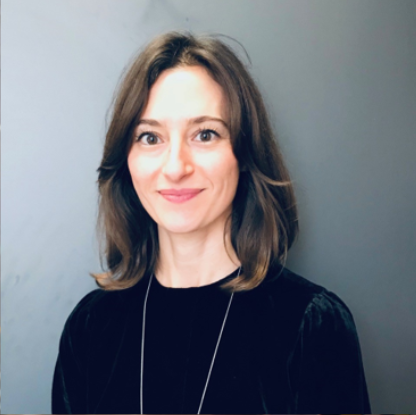}}]{Oya Celiktutan} received the Ph.D. degree in electrical and electronic engineering from Bogazici University, Turkey, in collaboration with the National Institute of Applied Sciences of Lyon, France, in 2013. She spent several years as a postdoctoral researcher with the Queen Mary University of London; the University of Cambridge; and Imperial College London, United Kingdom. 
Since 2018, she has been an Assistant Professor (Lecturer) with the Centre for Robotics Research, Department of Engineering, King’s College London, United Kingdom, where she is the Head of Social AI \& Robotics Laboratory. Her research focuses on computer vision and machine learning, applied to human behaviour understanding and generation, social signal processing, and human--robot interaction. 
\end{IEEEbiography}

\vskip -2\baselineskip plus -1fil

\begin{IEEEbiography}[{\includegraphics[width=1in,height=1in,clip,keepaspectratio]{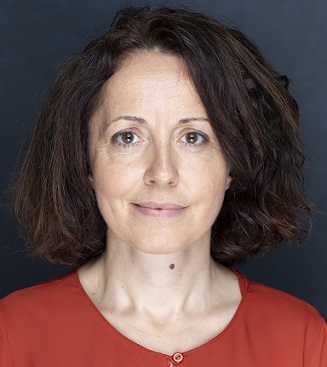}}]{Hatice Gunes}(Senior Member, IEEE) is a Professor of Affective Intelligence and Robotics (AFAR) at the Department of Computer Science and Technology, University of Cambridge, United Kingdom, leading the AFAR Lab. Her expertise is in the areas of affective computing and social signal processing cross-fertilizing research in human behaviour understanding, computer vision, signal processing, machine learning, and human--robot interaction. She has published over 125 papers in the above areas. Dr Gunes is the former President (2017-2019) of the Association for the Advancement of Affective Computing, was the General Co-Chair of ACII 2019, and Program Co-Chair of ACM/IEEE HRI 2020 and IEEE FG 2017. Her research has been supported by various competitive grants, with funding from the Engineering and Physical Sciences Research Council, UK (EPSRC), Innovate UK, British Council, Alan Turing Institute and EU Horizon 2020. She is currently a Fellow of the EPSRC and was previously a Faculty Fellow of the Alan Turing Institute.
\end{IEEEbiography}

\vskip -2\baselineskip plus -1fil

\begin{IEEEbiography}[{\includegraphics[width=1in,height=1in,clip,keepaspectratio]{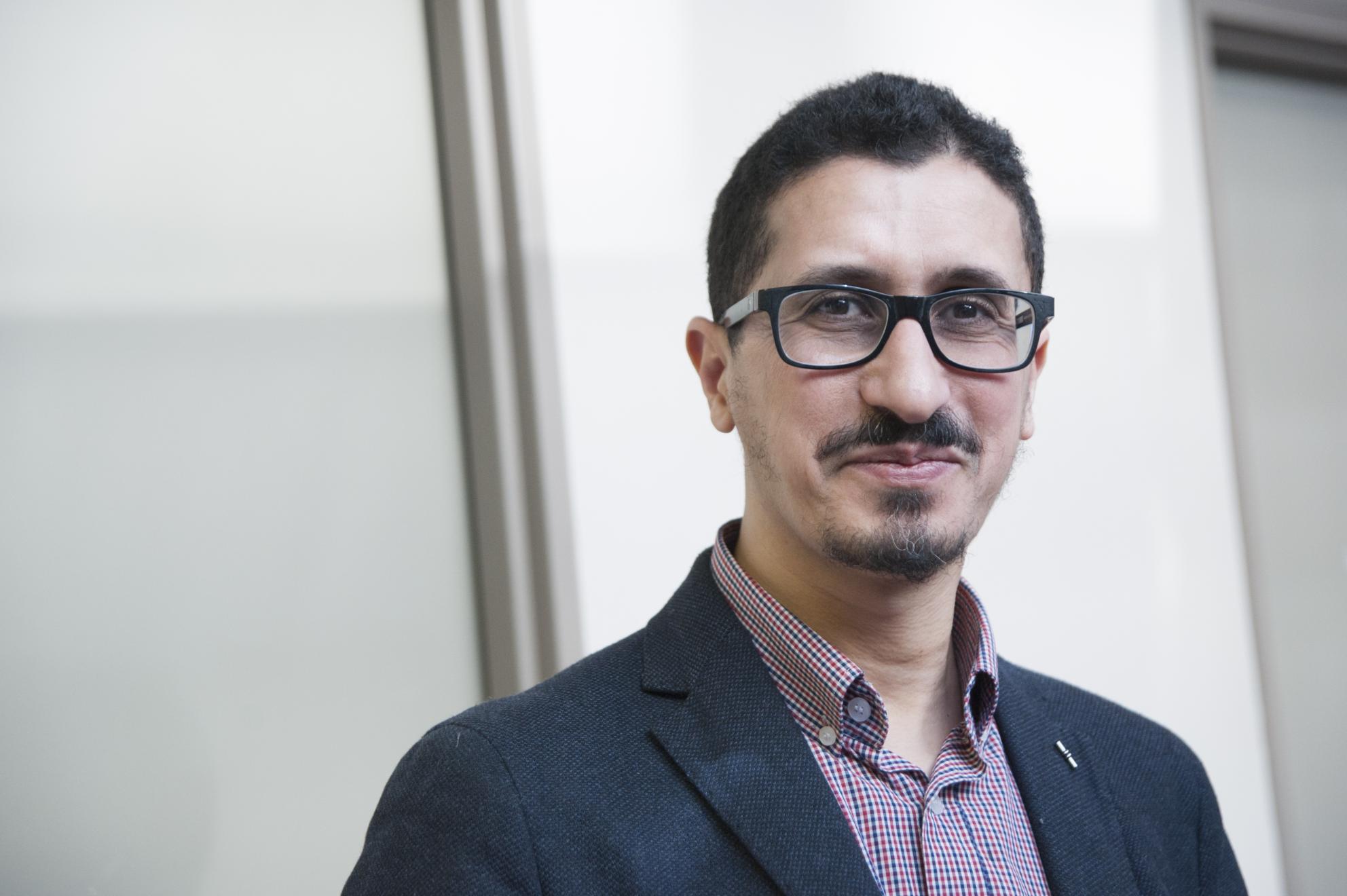}}]{Mohamed Chetouani}  is a Professor of signal processing and machine learning for human-machine interaction. He is affiliated to the PIRoS  research team at the Institute for Intelligent Systems and Robotics (CNRS UMR 7222), Sorbonne University. His activities cover social signal processing, social robotics and interactive machine learning with applications in psychiatry, psychology, social neuroscience and education. Since 2018, he is the coordinator of the ANIMATAS  H2020 Marie Sklodowska Curie European Training Network. Since 2019, he is the President of the Sorbonne University Ethical Committee. 
He is member of the management board of the International AI Doctoral Academy initiated by European networks of AI excellence centers. He is member of the EU Network of Human-Centered AI. He was Program co-chair of ACM ICMI 2020. He is General Chair of VIHAR 2021 and ACM ICMI 2023.
\end{IEEEbiography}
\end{document}